\documentstyle[prd,aps,preprint,epsfig,floats]{revtex}
\tightenlines

\input epsf.tex
\def\DESepsf(#1 width #2){\epsfxsize=#2 \epsfbox{#1}}

\newcommand{\be}{\begin{equation}}
\newcommand{\ee}{\end{equation}}
\newcommand{\bea}{\begin{eqnarray}}
\newcommand{\beas}{\begin{eqnarray*}}
\newcommand{\eea}{\end{eqnarray}}
\newcommand{\eeas}{\end{eqnarray*}}
\newcommand{\ba}{\begin{array}}
\newcommand{\ea}{\end{array}}
\newcommand{\bi}{\begin{itemize}}
\newcommand{\ei}{\end{itemize}}
\newcommand{\ben}{\begin{enumerate}}
\newcommand{\een}{\end{enumerate}}
\newcommand{\la}{\langle}
\newcommand{\ra}{\rangle}
\def\ls{\mathrel{\lower4pt\vbox{\lineskip=0pt\baselineskip=0pt
           \hbox{$<$}\hbox{$\sim$}}}}
\def\gs{\mathrel{\lower4pt\vbox{\lineskip=0pt\baselineskip=0pt
           \hbox{$>$}\hbox{$\sim$}}}}
\begin{document}

\draft

\preprint{\vbox{
\hbox{UMD-PP-03-065}
}}

\title{ Large Extra Dimensions and Decaying KK Recurrences }

\author{\bf R.N. Mohapatra$^1$, S. Nussinov$^2$\footnote{On sabatical
leave from Department of Physics, Tel-Aviv University, Israel.}
 and A. P\'erez-Lorenzana$^3$}

\address{$^1$ Department of Physics, University of Maryland, College Park,
MD20742, USA\\
$^2$ Department of Physics, University of South Carolina, Columbia,
SC-29208\\
$^3$ Departamento de F\'{\i}sica,
Centro de Investigaci\'on y de Estudios Avanzados del I.P.N.\\
Apdo. Post. 14-740, 07000, M\'exico, D.F., M\'exico}

\date{August, 2003}
\maketitle

\begin{abstract}
  We suggest the possibility that in ADD type brane-world
scenarios, the higher KK excitations of the graviton may decay to
lower ones owing to a breakdown of the conservation of extra dimensional
``momenta'' and study its implications for astrophysics and cosmology.
We give an explicit realization of this idea with
a bulk scalar field $\Phi$, whose nonzero KK modes acquire
vacuum expectation values.
This scenario helps to avoid constraints on large extra
dimensions that come from gamma ray flux bounds in the direction of
nearby supernovae as well as those coming from diffuse cosmological gamma
ray background. It also relaxes the very stringent limits on reheat
temperature of the universe in ADD models.
\end{abstract}

 \section{Introduction.}
For many decades, starting with the pioneering work of Kaluza and
Klein, the possibility that there may be extra hidden space
dimensions has been considered seriously primarily for reasons such
as the unification of gravity with other forces of nature. String
theories provided a very compelling reason for extra space
dimensions from purely theoretical considerations of conformal
invariance and Lorentz invariance. There was a mini-boom of activity
on the idea of extra space dimensions in the late 70's and early 80's
following this
realization.   It seemed natural to assume
that possible extra dimensions need not be as tiny as $l_{Planck}=
M_{P\ell}^{-1}\sim 10^{-33}$ cm, which, in many ways, is the smallest
distance possible, but rather any where in the region between this
and the direct experimental upper bound from colliders i.e. $l\sim
O(TeV^{-1})\sim 10^{-18}$~cm.

The recent mega-boom of theoretical interest in the subject came in
the wake of several papers which suggested the more radical
possibility that the hidden space dimensions could be
large~\cite{witten,early1}, even as large as a
millimeter~\cite{ADD}, which is far above the $10^{-18}$ cm
mentioned above without contradicting any known data. The class of
theories where this possibility arises is one
where the universe is assumed to have a brane-bulk picture, with
gravitons but {\it not} the standard model particles
propagating over a (much) larger extra dimension with $R= r_0 \times
10^{-3}$ cm\cite{ADD,early2,dbranes}. The gravitational coupling to
ordinary SM matter residing on the SM brane of size $l$ is then
diluted by a factor $f=[l/R]^d$ with $d$ the number of extra
dimensions. This ``naturally'' explains the weakness of gravity in
the absence of a high fundamental energy scale. Also in these theories
a crossover of the gravitational force law from the $1/r^2$ to
$1/r^{(2+d)}$ should occur for $r<R$, adding a strong motivation to
test Newton's law of gravity in an entirely new
domain~\cite{expgrav}. Choosing $d=2$ allows  satisfying the
G(Newton) constraint:
 \begin{eqnarray}
  \frac{l^2}{R}=l_{Planck}\sim 10^{-33}~~ cm
  \label{eq0}
 \end{eqnarray}
with $l\sim 10^{-18}$~cm consistent with Tevatron bounds and
$R\sim 10^{-3}$~cm consistent  with recent tests of the Newtonian
$1/r^2$ law down to $O(100)$ micron distances~\cite{expgrav}.

These models~\cite{ADD} have interesting phenomenology, including
enhanced neutrino cross-sections at high energy~\cite{ref2}, and the
possibility of producing microscopic Black Holes in accelerator and
cosmic ray physics~\cite{bhs}. Also with the various SM fermions
restricted  to specific  ``domain walls'' at locations $[v(i),u(i)]$
the masses mixing (CKM matrix elements) and potentially also the
phases of the latter may be ``explained''~\cite{ref3}.

The large extra dimensions in this picture coexist with a low TeV
level fundamental scale of nature which provides a new way to
resolve the gauge hierarchy problem (i.e. why is $M_Z \simeq
10^{-16}M_{P\ell}$ ?). An added bonus of these theories is the
possibility that string physics could be accessible to the
present generation of colliders~\cite{lykken1,lykken2,colliders,cdf}.

Despite these novel features, interest in such models has
declined for various reasons, one of which is the lack of a
convincing  theoretical framework explaining the stability of
the thick branes and the domain walls therein and another
is the failure to find any hints in favor of this idea
in gravity measurements\cite{expgrav} or in high energy collider
experiments~\cite{cdf}.

A third reason is perhaps  the drastic change in our thinking
about astrophysics and cosmology that is forced on us by large
extra dimensions approach. For example the presence of a closely
packed Kaluza Klein  (KK) tower of massive graviton modes that
is characteristic of these models, leads to a picture of early
universe which is somewhat murky leaving unclear the nature of
such issues as inflation~\cite{hdcosmo} and
baryogenesis\cite{baryo}. The most unusual implication of this
picture is that the universe cannot have a reheat temperature of
more than few MeV; otherwise, the heavy KK modes that are
produced overclose the Universe~\cite{ADD,han1}. A Universe
starting out at a temperature of a few MeV may barely have enough
room for a successful BBN but it poses quite a challenge to
understand the origin of matter~\cite{baryo}. Our goal is to
address these as well as related issues of large extra dimension
models (denoted as LED in this paper).

Strong upper bounds on $R$ or $l$ are derived from the following
astrophysical considerations~\cite{hannestad,various,moreastro}. The
emission of KK graviton recurrences from supernovae can
\begin{itemize}

\item (a) cause too quick a cooling of the latter and/or

\item (b) the subsequent
 decay of the emitted KK recurences into photons may generate too many
 diffuse Gamma rays.

\end{itemize}

In the conventional ADD picture, massive KK modes of the gravitons
only decay to photons, $e^+e^-$, etc., since their decays to lighter
graviton KK modes are forbidden by the conservation of extra
dimensional momentum. KK gravitons couple to brane matter with
Planck mass suppression~\cite{ADD,lykken2}, and therefore, lifetimes
for these decays are in the range of $\sim 10^{7}$ years for a KK
mode with mass of 100 MeV. This coupled with very close spacing of
the KK levels is the reason for strong constraints on the parameters
of the model as we review below.

In this paper, we propose modified ADD type models where
conservation of the extra dimensional momentum is violated
allowing the massive KK modes of the gravitons to decay to other
lighter graviton KK modes. There are then two competing decay
modes of the massive KK gravitons: the conventional one to
standard model particles such as electrons, neutrinos and
photons that are confined to the brane and new ones to lighter
KK modes of gravitons. For reasonable range of parameters, the
new modes i.e. $g_{KK}\rightarrow g_{KK,1}+g_{KK,2}$ can
dominate. In that case, it has the following kinds of effects:
(i) it can prevent in the simplest manner possible overclosure
of the universe by stable KK's relics, avoiding
 which otherwise requires some
pattern of expanding and shrinking dimensions and very late
inflation; (ii) it also can help to eliminate one class of
supernova bounds that arise from the diffuse gamma ray background
caused by the decays of the KK halo of neutron stars and
(iii) finally, it can also relax the bounds that arise from cosmic
diffuse gamma ray background arising from the decay of the graviton KK
modes produced in the early universe. It however
does not seem to affect the bounds that arise from supernova cooling by
emission of KK modes.

This paper is organized as follows: in sec. 2, we review in a
crude and intuitive manner how the cosmological bounds arise
in the LED models; in sec. 3, we discuss the way the astrophysical
bounds (a) and (b) above arise; in sec. 4 we review the argument
forbidding the inter KK decays in the case of flat manifold
compactification and note an amusing
generalization for all compact manifolds with
``KK Recurrences'' transforming as any representations of the
underlying ( compact ) Lie-algebra. In sec. 5, we show how the
spontaneous breakdown of extra dimensional momentum conservation
allows the gravitational decays of the higher KK modes of the graviton and
estimate the new decay rates for various scenarios for the breakdown of
the extra dimensional momentum;
 in section 6, we show how this new decay mode affects the astrophysical
and the cosmological bounds.
 Finally,  section 7 is devoted to discuss how  the KK modes
of an scalar field may pick up non zero vacuum expectation
values. In particular, we show how  a single scalar KK mode with
nonzero extra dimensional momentum induces nonzero vev for a
whole tower of modes, leading to what we call a ``condensate
tower''. In section 8, we conclude with a summary of the results.

\section{Cosmological bounds on the size of extra dimensions}

Any primordial abundance of KK gravitons recurrences can be
diluted by inflation. However, as noted in~\cite{ADD,han1}
subsequent reheating even to temperatures of a few MeV will
regenerate for extra dimension size of a millimeter a density of
essentially stable KK modes of the graviton with masses $\leq
T_R$, where $T_R$ is the reheat temperature.

In order to see very crudely how the cosmological constraints arise,
let us make the most conservative estimate for the emission rate of
the $g_{KK}$ modes from the collisions of the standard model
particles during the Hubble time $dt_H\sim
-2dT\frac{M_{P\ell}}{\sqrt{g_*}T^3}$. Typical rates for $g_{KK}$
emission from these collisions for each KK mode are:
\begin{eqnarray}
R_{g_{KK}}~\sim \frac{T^3}{M^2_{P\ell}}.
\end{eqnarray}
But at a given temperature, all KK modes upto $T$ are produced.
Thus in the Hubble time $dt_H$, the total number of $g_{KK}$
modes can be estimated to be
\begin{eqnarray}
 N_{g_{KK}}\sim \int dt_H R_{g_{KK}} (TR)^2 \sim
 \frac{2}{3} \frac{T_R^3R^2}{\sqrt{g_*}M_{P\ell}}  =
 \frac{2}{3}\frac{M_{P\ell}~T^3_R}{\sqrt{g_*}M^4_*}.
\end{eqnarray}
Note that most of the particles have masses close to the reheat
temperature since the production rate goes down very sharply with
temperature.
Since these particles are stable, they survive all the way to today and
therefore their number density today is roughly $\sim N_{g_{KK}} T^3_0$.
Their contribution to energy density today can be estimated to be
\begin{eqnarray}
\rho_0\simeq \frac{2}{3}\frac{M_{P\ell}~T^3_0 T^4_R}{\sqrt{g_*}M^4_*}
 \end{eqnarray}
since modes close to $T_R$ dominate as already mentioned. Setting
$\rho \leq \rho_{cr}$, we get $T_R \sim 1$~MeV~\cite{ADD,han1}. We
have summarized the essential points in this argument. For a more
detailed analysis and the resulting refined bound, we refer to the
paper \cite{han1}. We have neglected purely gravitational mechanisms
for $g_{KK}$ production, which would only enhance the $N_{g_{KK}}$
and strengthen the bound.

 \section{The astrophysical bounds}

We first note that for the case of two extra dimensions, the
number of KK modes with masses $ < m$ strongly increases with $m$,
\begin{eqnarray}
  n_{(KK;~ m(KK)\, <\, m)}\sim  [m\cdot R]^2\sim r^2_0
  {[m/MeV]}^2 \cdot 2.5 \times 10^{15}~.
\end{eqnarray}
In cosmological or astrophysical settings where the temperature
exceeds $m$, all these modes will be excited. We can now see the
origin of the astrophysical bounds due (a) to KK emission and (b) to
possible subsequent electromagnetic decays of the latter  as
follows:

 (a) During the first few seconds of supernova collapse, the temperature
of the collapsing core has a temperature $T=y\times 10$~MeV with
$y\sim 1-5$. The spectrum of neutrinos from  SN 1987a indicates the
O(5) MeV temperature of the more extended, cooler, ``neutrino
sphere".
 To simplify we assume that roughly
 $N\simeq (RT)^2={(r_0\cdot y)^2}\cdot (2.5)\times 10^{17}$
 KK 's with  masses $< 30$~MeV are emitted at the same rate as massless
gravitons. They are emitted
from relativistic particles inside the SN at a rate
 \begin{eqnarray}
 (\tau)^{-1}\simeq T^3/M_{P\ell}^2
    \simeq (1.6)\cdot y^3~10^{-18}~ {\rm sec}^{-1}~.
 \label{eq1}
 \end{eqnarray}
 Hence the  energy of such particle ( neutrinos, electrons/ positrons or
photons ) dissipates N times faster i.e during  :
 \begin{eqnarray}
  \Delta t\simeq
   \tau/N\simeq 2.5~{\rm sec} \cdot [r_0]^{-2}\cdot [y]^{-5}~.
 \end{eqnarray}
For $r_0\geq 0.1$ and $y\simeq 3$,  the emission of KK's yields
cooling times $\Delta t\sim 10$ sec, commensurate with the
observed duration of the SN 1987a neutrino pulses.
 Hence $R> 10^{-4}$ cm is excluded as the observed pulses would
be shorter and weaker.
 Careful calculations by various groups~\cite{various} yield $R\leq
0.7\times 10^{-4}$ cm.
  This bound relies only on the emission of the KK recurrences, not on
possible radiative decays of the later.

 (b) Turning now to the second class of supernova
bounds~\cite{hannestad}, it arises as follows: for $r_0< O(1)$ only
a fraction ($<30$\%) of the $ 10^{53}$ erg  gravitational collapse
energy in supernovas is emitted via KK recurrences. The mildly
relativistic KK 's towards the upper end of the mass spectrum of
those emitted from the supernova survive the cooling process thanks
to their enhanced number and form a halo around the remnant neutron
star. They decay at the same rate as in Eq.~(\ref{eq1}) with
$T\simeq E\simeq m\simeq y\times 10$ MeV.
 Taking then $y\sim 3$, as above, we find a radiative KK decay rate:
${\tau (R)}^{-1}\sim 8\times 10^{-18}$~sec$^{-1}$ which is
independent of the size of the extra dimensions.
 If the SN is at a distance of k kilo-Parsecs, the fraction of KK's
 decaying en-route is
 \begin{eqnarray}
    f\simeq 0.4~k\times 10^{-7}. \label{eq5}
 \end{eqnarray}

Unlike for putative decays of light extreme\cite{goodman} -relativistic
neutrinos, photon from decaying mildly  relativistic KK's will
not appear shortly after the  neutrino bursts  and will not come
precisely from the Supernova's direction. Thus the strict limits
on Gamma/X-rays appearing shortly after the neutrino burst used
to strongly limit electromagnetic decays of neutrinos are not
useful here.

However Raffelt and Hannestad~\cite{hannestad} have correctly
noted that past supernovae collectively produce a ``diffuse
gamma ray background'' at E~$\sim 30$~MeV exceeding experimental
upper bounds- even if only a minute fraction $\sim 10^{-5}$ of
the collapse energy is emitted via KK's.

 The following crude estimates are no substitute for the careful,
original work, yet may be of some interest for an overall
understanding of the reason for the limit. The total observed
electromagnetic ``diffuse" energy in a $\sim 10$~MeV interval
around 30 MeV that can 
 $\simeq ~10^{-8}$~eV/cm$^3$ which is roughly $ 10^{-10}$ times the
baryonic matter density which constitutes~1\% of the critical
density\footnote{This bound gets further strengthened to about $\leq
10^{-9}$ eV/cm$^3$ if one uses the bound deduced\cite{esp} from
the EGRET data~\cite{egret} observations and extrapolate it according
to the $E^{-2}$ law.}. 
 To satisfy the bounds on the diffuse 30 MeV gammas only
$10^{-10}$ (!) of the baryonic rest mass can convert into such
photons via emission of KK's with masses $\sim 30$~MeV from SN's
and the subsequent photonic decay of the latter.

 To simplify and minimize red-shift and evolution effects consider only
supernovae at less than $10^6$ Kiloparsecs. From Eq.~(\ref{eq5}), we
find that $\sim$~4\% of the KK's decay over this distance; hence the
total energy  emitted as  ``heavy'' i.e $\sim 30$~MeV KK's should be
less than $(2.5)\times 10^{-9}$  of the baryonic rest mass.

Assume that type II supernovae occurs at $\sim 1/30$ years per
average galaxy with mass  $\sim~10^{12}~m_\odot$, in the above
$z<1/3$ range i.e $10^{-4}$ of all stars  collapse over the
lifetime of the Universe. In each collapse, approximately
$\sim~.25~m_{\odot}$ of gravitational energy is emitted out of
which about $\sim ~.1~m_\odot$ can be emitted in the form of
``massive'' KK's without violating the first ``cooling'' bound.
 Thus $10^{-5}\times 0.04=4\times 10^{-7}$ of the baryons rest mass
could appear as 30 MeV gammas !
 This exceeds the upper bound of $10^{-10}$  by 3.5 orders of magnitude
unless the bulk dimension
 \begin{eqnarray}
  R< 1.6 10^{-5}~~~ {\rm cm}
 \label{eq6}\end{eqnarray}
which is 70 times smaller than the first bound.

 In any Type II supernova we expect $\sim .6~m_\odot$ of iron to be
ejected i.e six times more than the energy in massive KK's. The
cosmological iron abundance of $10^{-3}$ of baryonic density thus
implies  that $1.6 \times 10^{-4}$ ( 16 times more than in the
previous estimate !) of the baryonic rest mass converts to massive
KK's in all supernovae collapses. This requires
 $R<4\times 10^{-6}$~cm
which is four times more restrictive than (\ref{eq6}).

  The above bounds push the value of the size of the extra dimension $R$
far below what may be seen in direct
gravitational measurement searches.
 Also Eq.~(\ref{eq0}) implies then that the thickness of the standard
model brane is bound by
 \begin{eqnarray}
   l< 10^{-19}~{\rm cm} \simeq (100~{\rm TeV})^{-1}~;
 \end{eqnarray}
which is beyond the LHC range.

 \section{Forbiddenness of intra-KK decays in the bulk and
generalizations.}

A basic rule of quantum mechanics is that corresponding to every
noncompact space-like direction, there is a conserved quantum number
which is familiar to us as momentum. When space is compactified (like
the case of a particle in a box), the values of the momenta become
discrete but the conservation law still holds. In the case of compact
extra space dimensions, processes involving particles that live in the
higher dimensions must conserve the corresponding discrete momentum
quantum number. For example, in an allowed process
$KK(k_1)+KK(k_2)\rightarrow KK(k_3)+KK(k_4)$, we must satisfy the
condition $k_1+k_2=k_3+k_4$. Note that conservation of extra
dimensional momentum forbids all inter KK decays. This can be seen for
the case of two dimensions as follows: suppose a particle of extra-D
momentum $\vec{n}$ decays to two particle with extra momentum $(\vec{n}_1,
\vec{n}_2)$. Then extra-D momentum conservation implies that
$\vec{n}=\vec{n}_1+\vec{n}_2$. The mass of the decaying particle, if it
has no bare mass term, is given by 
 $m_{KK}() =
 \sqrt{ (\vec{n}_1 +\vec{n}_2)^2 }\cdot R^{-1}$.
 Conservation of extra dimension momentum otgether with that of ordinary
momentum then implies that the
initial decaying mass equals the sum of the final masses leaving
a vanishing ( ordinary 3 dim. ) phase space for the decay. The decay 
is therefore kinematically forbidden.

If the compactification is curved rather than flat, one can
define analogs of extra dimensional momenta (called charges) and
only bulk transitions conserving these charges will be allowed.
Amusingly enough, this persists in even more restrictive form in
the general case when the underlying internal group is any
semi-simple compact Lie group and the extra dimensions manifold
are any coset space.
 The KK particles are then invariantly
characterized in the extra dimensions by a representation $R$ of
the group $G$ to which they belong.
 The mass corresponding to the (internal) compact part of the
Laplacian will be proportional to $C_{2}(R)$ the Quadratic Casimir
operator for the representation.
  The decay $A\rightarrow B+C$ with $A$, $B$ and $C$ in representations
$R(A)$, $R(B)$ and $R(C)$ can occur only if the direct product
$R(C)\otimes R(B)=\sum_i R(i)$ includes $R(A)$.

  One can then show that $C_{2}(C) + C_{2}(B) > C_{2}(A)$
implying that $m(A) < m(B) + m(C)$ and any allowed intra KK decay
will always be below threshold\footnote{We will not reproduce
here the simple proof of this relation which has been kindly
provided to us by Prof. Bernstein of the Math. Dept. in Tel-Aviv
Univ.
 It utilizes the Harish-Chandra formula for $C_{2}(R)$
in terms of the maximal weights of $R$ and simple features of
such weights such as the additivity upon direct product.}.
 This can be motivated by verifying it as an algebraic identity for all
representations of $SU(2)$:
 \begin{eqnarray}
  \left(j.(j+1)\right)^{1/2} + \left(k.(k+1)\right)^{1/2}
  > \left((j+k)+j+k+1\right)^{1/2}~;
 \label{ineq}
 \end{eqnarray}
 and also note the similarity to the triangular inequality which holds for
convex internal spaces (or compact Lie groups).

 \section{Spontaneous Breaking of the  translational invariance in the
compact dimensions and decay of KK gravitons}

In this section, we propose a model where the graviton KK modes are
unstable and appear to provide a way around some of the
astrophysical and cosmological constraints. Our proposal is to break
the translation invariance along the hidden spacelike fifth and
sixth directions while keeping in tact both Lorentz and translation
invariance in the familiar 3+1 space time. In such a theory, the
inequality in
Eq.~(\ref{ineq}) will not be obeyed and therefore the heavier KK
modes of gravitons can decay. One could in principle extend this
also to extra dimensions with curvature; but here we focus only on
flat extra dimensions in the spirit of the ADD model where the
presence of such a spontaneous breaking, which leads to breakdown of the
KK momentum conservation equation $\vec{n}_A+\vec{n}_B = \vec{n}_C$
in a particular reaction. As a result for any KK mode of the graviton, 
decays such as $A\rightarrow B + C$ are no more kinematically forbidden.

To implement our proposal, we take the usual brane-world scenarios where
the standard model particles and forces are all confined to a 3-brane
and gravity is in the bulk. We add a new scalar field
$\phi(x,\vec{y})$ in the bulk; here $x$ stands for the familiar
space-time and $\vec{y}=(y_1,y_2)$ denotes the two hidden space
dimensions. This bulk field can be expanded in terms of its Fourier
components, which are fields that depend only on $x$ and are the
effective fields on the brane:
\begin{eqnarray}
 \Phi(x,y)~=~\sum_{n_1,n_2}~\frac{1}{2\pi\, R}~\phi_{\vec{n}}(x)\,
            e^{i\, \vec{n}\cdot\vec{y}/R}~;
\end{eqnarray}
for $\vec{n} = (n_1,n_2)$. We then consider an appropriate Higgs
potential for the bulk field such that KK modes of $\phi$, upto a
large number, acquire vacuum expectation values, 
 \begin{eqnarray}
 \la\phi_{\vec{n}}\ra~=~v_{\vec{n}}~.
 \end{eqnarray}
This can be done for example by choosing a
negative mass term for the $\phi$ field.
We defer the discussion of this to a subsequent section, where we give
some explicit examples. Note that as mentioned earlier, this leaves the
four dimensional Lorentz as well as translation symmetry unbroken.

Let us next postulate the existence of a coupling of the form
 \begin{eqnarray}
 {\cal L}_I~=~
 \lambda\, M^4_*\, \Phi\, g_{AB}\, g_{CD}\, g_{EF}\, {\cal C}^{ABCDEF}~.
 \label{phiggg}
 \end{eqnarray}
This kind of  $g^3$ coupling could arise from generalized
coordinate transformation invariant higher dimensional
interactions of the form $\int d^4x\, d^2y\,\sqrt{-G}\,
T^A_A(x,y)$ after expansion with $g_{AB}=
\eta_{AB}+M^{-2}_*h_{AB}$ upto third order in $h$, and with  $T$
as  the vacuum energy. We will use, however, Eq.~(\ref{phiggg})
as a toy model interaction for our analysis below.

At the level of the  effective four dimensional theory we get the
following couplings among KK modes:
 \begin{eqnarray}
 {\cal L}_I~=~
 \lambda\, \frac{M^2_*}{M^2_{P\ell}}\,
 \phi_{\vec{n}_\phi}\, h_{\vec{n_1}}\,
 h_{\vec{n_2}}\, h_{\vec{n_3}}\, {\cal C}~.
 \end{eqnarray}
To see how one gets the coupling strength of order
$\left(\frac{M_*}{M_{P\ell}}\right)^2$, note that while expanding the
$\Phi$ field in terms of 4-D fields, we get a factor $\frac{1}{R}$ and we
have $g_{AB}= \eta_{AB}+\frac{1}{M^2_*R}h_{AB}$. Noting that integration
over $d^2y$ gives $R^2$ and using $M^4_*R^2=M^2_{P\ell}$, we get the
strength of the interaction given in Eq. (15).
 Clearly in this equation we have $\vec{n}_\phi+\sum_i
\vec{n}_i=0$. Once the four dimensional scalar field acquire a
vacuum,  $\la\phi_{\vec{n}_\phi}\ra\neq 0$, this interaction
leads to $g_{KK}\rightarrow g_{KK}+g_{KK}$ as noted before, with
proper identification of $h$ as $g_{KK}$.

We show in sec. 7 that once one $\vec{n}_\phi\neq 0$ mode of $\Phi$ has
nonzero
vev, a tower of $\phi_{\vec{n}}$'s acquire  nonzero vev by induction. We
will call this a ``condensate tower''. The condensate tower plays an
important role in enhancing the decay rate of the higher KK modes. To
calculate the decay rate of a high graviton mode
$g_{KK}$, we first note that for each $\la\phi_{\vec{n}_\phi}\ra\neq 0$,
there are approximately $(\mu_{KK}R)^2$ open channels which counts
 the number of ways
the equation $\vec{n}_i-\vec{n}_{f1}-\vec{n}_{f2}= \vec{n}_{\phi}$ can be
satisfied for a given $\vec{n}_{\phi}$, where $\vec{n}_{i,f}$ denote the
quantum 
numbers of the initial and final state KK modes.
Then there is the contribution from
each of the different $\la\phi_{\vec{n}_\phi}\ra\neq 0$ in the condensate
tower contributing to
the decay. Since each condensate tower consists of approximately
$(\mu_{KK}R)^2$ entries, this gives an extra factor $(\mu_{KK}R)^2$ in the
decay rate (assuming of course that the vevs are of same order.
Combining all these, we get
 \begin{eqnarray}
  \Gamma^{TOT}_{\vec{n},~ g\rightarrow g g }~\simeq
  \frac{\lambda^2\mu_{KK}\sum_{\vec{n}}\la\phi_{\vec{n}}\ra^2}{2\pi
M^2_{P\ell}}~,
 \end{eqnarray}
 and if all vevs are assumed to be same, we get
\begin{eqnarray}
  \Gamma^{TOT}_{\vec{n},~ g\rightarrow g g }~\simeq
  \frac{\lambda^2\mu^3_{KK}R^2\la\phi_{\vec{n}}\ra^2}{2\pi
M^2_{P\ell}}~= \frac{\lambda^2\mu^3_{KK}\la\phi_{\vec{n}}\ra^2}{2\pi 
M^4_*}
 \end{eqnarray}
 Compare this with the
 rate for a  KK graviton decaying into photons which goes as
$\Gamma_{g\rightarrow \gamma\gamma}\simeq \mu_{KK}^3/M_{P\ell}^2$.
It is clear that for a large enough vev, the inter KK decay
channel dominates over the decay into standard model
particles, since
 \be
 \Gamma_{\vec{n},~g\rightarrow g g}^{\rm total} \simeq
 \Gamma_{g\rightarrow \gamma\gamma}
 \left({\la\phi_{\vec{n}}\ra M_{P\ell} \over M^2_*}\right)^2 ~.
 \label{gamman}
 \ee
For instance, if $\langle\phi_{\vec{n}}\rangle\sim M_*\sim $ TeV, we get
 an enhancement by at least a factor of  $10^{30}$. Thus, the KK graviton will
predominantly decay into lighter KK gravitons.

\section{Implications of the modified ADD models}

\noindent{\bf Neutron star/supenovae limits}
We note from 
Eq.~(\ref{gamman}) that for reasonable values of $\langle
\phi_{\vec{n}}\rangle$ of order of a TeV, only a tiny fraction of the
$\sim 30$~MeV mass KK gravitons produced in old supernova could latter
decay into photons and contribute to the diffuse gamma ray
background. Naivlely taking the branching ratio in Eq.~(\ref{gamman}), one
can see
that at most a fraction $f\simeq 10^{-10}$  of the expelled KK modes would
decay into visible channels. As noted in Ref.~\cite{hannestad}, the lower
bound on the value of the fundamental scale $M_*$ of LED models goes like
$f^{1/4}$. In the absence of the new inter KK decay modes, $f\simeq 1$.
Now that the fraction decaying to gamma rays is reduced, the lower bound
on $M_*$ is also reduced by $10^{-2.5}$ leading to $M_*\sim 1$ TeV.

Moreover, for EGRET's direct observation towards nearby neutron
stars, we should  notice that one  has to take into consideration
that the life time of a 30 MeV KK graviton would be much shorter
than the one estimated in the usual case with only visible decay
channels. Now one rather gets $\tau\ls 10^8$~sec for the above
choice of parameters (the value considered for the bounds in
\cite{hannestad} was
about $10^{18}$~sec). This drastically reduces the present day
source strength of MeV gamma rays by a factor of
$exp(-t_{NS}/\tau)$, where
 $t_{NS}\sim 10^6$~yrs is a typical age of a  neutron star.
Thus, the present day source density of massive graviton KK modes should
be many many orders of magnitude smaller than in the case with
conservation of extra dimensional momenta. This
makes the contribution to gamma emission by the nearby neutron
stars negligible even for the closest such objects. This
implies that the bounds on the size of extra dimensions obtained
from the direct observations from supernova remnants derived in
Ref.~\cite{hannestad} do not apply.

\vskip1em

\noindent{\bf Neutron star heating}

It was suggested in Ref.~\cite{hannestad}, that the halo of
heavy KK modes that should surrounds a neutron star remnant
could heat up the star by the photons, $e^+e^-$ pairs, and other
decay products of the KK gravitons that continuously hit the
star. In our model, with the dominance of the inter KK decay
channels, most of the heavy KK gravitons would decay into
lighter (relativistic) KK gravitons, which would then escape
from the star, thus, giving no contribution to the heating
process.

\vskip1em

\noindent{\bf Cosmology bounds}

As far as the upper bound on reheat temperature from cosmology
is concerned, the situation is also relaxed. The analysis is
somewhat more involved than the case of diffuse gamma ray
background. Here we briefly summarize the salient points that
make a difference in this analysis.

The production rate for the gravitational KK mode with mass
$\mu_{KK}$ at a temperature $T$ is roughly given by
$R_{G_{KK}}\approx \frac{T^3}{M^2_{P\ell}}$. In previous
discussions of this problem, the modes were assumed to be stable.
Since in our framework, the KK modes are unstable, we have to see
what their lifetimes are. As mentioned before, when all KK modes
of the $\phi$ field acquire similar vevs (we will show later this
to be the case), the decay width of a mode with mass $\mu_{KK}$ is
given by:
 \begin{eqnarray}
  \Gamma^{tot}_{g_{KK}}\simeq
  \frac{\mu^3_{KK}{\la\phi_{\vec{n}}\ra}^2}{2 \pi M^4_*}~.
 \end{eqnarray}
For $\mu_{KK}\simeq $ MeV and $M_*\simeq $ TeV, we get the life time of
such modes to be of order $10^{-9}$ sec. which is far less than the
age of the universe at 1 MeV. In fact the decay rate exceeds the Hubble
expansion rate at any temperature above this. What this means is that
above one MeV (the temperature at which BBN takes place), all heavy KK
modes produced decay instantaneously to the lowest KK modes. These modes
will redshift as the universe cools to the present temperature leading to
their contribution to present energy density $\rho_0$ to be:
 \begin{eqnarray}
 \rho_0\simeq
 \frac{2}{3}\frac{M_{P\ell}~T^3_RT^4_0}{\sqrt{g_*}M^4_*}~;
 \end{eqnarray}
where $T_0$ is the present temperature of the universe. Comparing
this with the expression in Eq. (4), we see that it implies that
$T_R\leq 10 $ MeV. Recall that a similar crude calculation without the KK
decay led to a value of $T_R\simeq 1$ MeV. Thus we are able to relax the
bound on the reheat temperature.

 This calculation is very crude and
is meant to illustrate the point. We expect that after the calculation is
done more carefully, the value of $T_R$ is likely to increase
somewhat. It is also interesting to note that the reheat
temperature $T_R$ scales like $(M_*)^{4/3}$; thus an increase in
$M_*$ by a factor of $\approx 5$ will increase the upper limit on
$T_R$ to a 100 MeV.

\vskip1em

We also note that the KK recurrence decay mechanism discussed here does
not seem to
affect the constraints on large extra dimensions from supernova emission 
effects.

\section{Inducing vevs for scalar KK modes}

There are many ways  a KK mode of a scalar field, $\phi$, could pick
up  a nontrivial vacuum expectation value. Here we would like to
comment some possibilities.

\bigskip
\noindent{\bf \it $\lambda \phi^4$ couplings}.- 

\bigskip

The simplest way to give
a vev to the scalar field, in 6D, is to have a potential of the form
 \be
  V(\phi) = - {1\over 2} \mu^2 \phi^2 + {\lambda\over 4 M_*^2}
  \phi^4~.
  \label{phi4}
 \ee
with $\mu^2 > 0$.
There is always a simple solution to above equation,
which is a y-independent vacuum (a true zero mode), given by the
minimum of the potential $\partial V(\phi)/ \partial \phi =0$
that is  given by
 $\la\phi\ra =\mu M/\sqrt{\lambda}$.
As already mentioned,  in the effective four
dimensional theory, the above vev is  larger, since
 \be
 \la\phi_0\ra=\la\phi\ra R =(MR) {\mu\over\sqrt{\lambda}}
 = \left({M_{P\ell}\over M_*}\right){\mu\over\sqrt{\lambda}} ~.
 \ee
Therefore, for an order one $\lambda$,  $\la\phi_0\ra$ can
naturally be many orders of magnitude larger than $\mu$. For our
purpose, we will choose $\mu$ of order $1/R$ or less so that we get a vev
only for the zero mode and a vev of
order $M_*$. Physically, this is a consequence of the highly
suppressed $\phi^4$ effective coupling in 4D.  Such a vev,
however, being only for the zero mode, does not break translational
invariance in extra dimensions and is therefore not useful for
the new phenomena we discuss.

There are however  other vacuum
configurations that are allowed for the potential, where higher KK modes
do get vevs. The configuration of such  vacua is in
general  highly non trivial, at least from the point of view of
the KK modes, as a perturbative analysis of the potential can
show. To simplify the arguments let us just consider a simplest 5D
model with a similar potential. The generalization to 6D should
is straightforward. By integrating out the extra
dimension in the action, one gets the effective potential in 4-D to be
 \be
 V(\phi_n) = -{1\over 2} \mu^2 \phi_0^2 -
 \sum_{n=1}^\infty \mu_n^2  \phi_n\phi_{-n} +
 {\tilde{\lambda}\over 4}\sum_{k,\ell,m,n=-\infty}^\infty
     \delta_{k+\ell+n+m,0}\, \phi_k\phi_\ell\phi_n\phi_m ~;
 \label{vphin}
 \ee
where $\mu_n^2 = \mu^2 - n^2\mu_c^2$, with $\mu_c=R^{-1}$ the
compactification scale, and $\tilde{\lambda} = \lambda/MR$.

Now, we notice that depending on the value of $\mu^2$, there can be a set
of modes for which the mass
term, $-\mu_n^2$, is  negative This will happen for modes starting with
the zero mode and continuing all
the way up to $n \simeq\mu R$. At a first glance, motivated by
what happens in four dimensional theories,   one could be tempted
to believe that only modes in that particular set will pick up 
vevs. However, although it is correct that there will be solutions
for the vacua with non zero KK vevs upto mode $\mu R$, it so happens that
such
configurations induce an infinite number of other modes
to have nonzero vev, even though they may have their $\mu^2_n >0$,
 as long as it gives a vev to a mode with nonzero extra-D
momentum.

 To clarify our claim, let us assume, for instance,
 that all modes but $\phi_{\pm n}$ are zero. Next we observe that
in the potential there is a coupling of the form
 $\phi_{\pm 3n}\phi_{\mp n}^3$ which contributes nontrivially to the
 minimization condition for $\phi_{\pm 3n}$. Indeed one gets
$\partial V/\partial\phi_{\pm 3n} \propto (\phi_{\mp n})^3\neq 0$.
Hence, one can conclude that $\phi_{\pm 3n}\neq 0$.
Once the mode $\phi_{3n}$ has nonzero vev, via the coupling,
 $\phi_{\mp 5n}\phi_{\pm3n}{\phi_{\pm n}}^2$,
which contributes  to  the minimization condition for
 $\phi_{\pm 5n}$, this will induce a vev for $\phi_{\pm 5n}\neq 0$.
 In a similar way one can proof that all $\phi_{(2k+1)n}$,
for every $k$ will have nonzero vev, just by noticing that there
are always some couplings  of the form
 $\phi_{(2k+1)n}\phi_{(2\ell +1)n}\phi_{(2m +1)n} \phi_{-[2(k+\ell +m)+3]n}$,
involving all non-null vev modes. Therefore, barring miraculous
cancellation happens among all those contributions, one is led to
the conclusion that typically the vacuum will have an infinite number of
modes  picking up a vev, giving rise to the condensate tower. Moreover,
one can see using this argument
that three classes of vacua are in general possible: (i) those containing
only modes with odd indices; (ii) those containing only modes with  even
indices; or (iii) with both classes of  modes having non zero vevs.

By going to six dimension our conclusion will not change except that we
must now include a second independent
index, which follows the same rules of conservation
that controls the way the mixing happens. The nonzero vevs in this case
will form a lattice.

One can also show that all these new vevs correspond to physical minima 
i.e. the second derivative are positive. This condition for
the minimum  reads
 \be
 {\partial^2 V\over \partial\phi_n\partial\phi_{-n}} =
   -\mu_n^2 + 3\tilde\lambda\left[ \phi_0^2 +
      2 \sum_{k=1}^\infty \phi_k\phi_{-k}\right]~.
 \ee
 As the term between brackets is common to all values of $n$,
one has that only the very first equation for the zero mode,
$n=0$, would be relevant to insure that any solution is a real
minimum. Moreover, from the same expressions we see that the
splitting among the physical masses of the KK modes would be
still given by the compactification scale, $\mu_c$.

\bigskip
\noindent{\bf \it Nambu-Goldstone modes}:

\bigskip

An immediate implication of spontaneous breaking of the extra dimensional
momentum is that it leads to the existence of massless scalar modes. This
can be illustrated using a one dimensional example. Note that the
Lagrangian involving the modes of the scalar field is invariant under the
transformation $\phi_n\rightarrow e^{in\theta}\phi_n$. As a result when 
$\phi_n$ acquires a nonzero vev, there is a massless Nambu-Goldstone mode
$\chi$. To see the expression for $\chi$, let us take the five dimensional
case and express the field
$\phi_n=\rho_n e^{i\chi_n/v_n}$. One then has:
\begin{eqnarray}
\chi ~=~ N_\chi [\sum_n n v_n \chi_n]
\end{eqnarray}
as the massless state. Here $N^{-1}_{\chi}~=~\sqrt{\sum_n n^2v^2_n}$. This
has
an obvious
generalization to the case of two dimensions.

If the scalar field is coupled to brane fields, then this global symmetry
is broken and the massless mode is a pseudo-Goldstone boson. The loop
corrections will induce a mass to this mode of order $m^2_\chi\simeq
\frac{h^2}{16\pi^2}M^2_*$

In either case, all our phenomenological considerations remain unchanged.

\bigskip
\noindent{\bf \it Bulk  Topological Defects.-} 

\bigskip
Strictly speaking,  a
y-dependent vacuum configuration for the $\phi^4$ potential is an
exact (non perturbative) solution to the equation of motion,
obtained directly from the six dimensional action, and having no
dependence on our standard four dimensions, thus, it is given by
the solution to
 \be
 {\nabla_\bot}^2 \phi + \mu^2 \phi = {\lambda\over 4 M_*^2}
 \phi^3~;
 \ee
where $\nabla_\bot = \partial_{y_1}^2 + \partial_{y_2}^2$.
 A particularly interesting  solution  is the
well known case of  domain walls configuration,
which can be written as
 \be
 \phi(y) = {v}~\tanh( \vec{\alpha}\cdot \vec{y})~;
 \ee
with $2\alpha^2 = \mu^2$ and $v = \mu M_*/\sqrt{\lambda}$. There
are, as is well known,  other non trivial solutions of this
defect-type. It is  clear, from the Fourier expansion, that in the
present case  all KK modes  will get a nontrivial vev, since
 \[
    \langle \phi_n\rangle
 = {1\over 2\pi\, R}\int_0^{2\pi\, R}\int_0^{2\pi\, R}\! d^2y\, \phi(y)\,
        e^{- i\vec{n}\cdot\vec{y}/R}
 = vR
     \left[{1\over 2\pi}
     \int_0^{2\pi}\int_0^{2\pi}\! d^2x \tanh(R\vec{\alpha}\cdot\vec{x})\,
        e^{2\pi i (\vec{n}\cdot\vec{x})}\,\right]
    \neq 0~.
 \]
Notice that the last expression within brackets would be just a
number for a given $n$. Thus, the overall value of the KK vevs
would  be just $M_*R\, \mu/\sqrt{\lambda}$.

\bigskip
\noindent{\bf\it Using hidden branes}.-

\bigskip

 Another possibility is to have a
vacuum on
some (hidden) brane
 shining into the bulk by generating  nontrivial vevs for all
KK modes~\cite{shiny} of the bulk scalar. The mechanism works whenever the
bulk
scalar field mixes with a brane field, $\chi$, that has a vev, for
instance through the coupling
 \be
- M_*\phi(y)\,\chi\,\delta({\vec{y}})~;
 \ee
which introduces a point-like source for the  $\phi$ vacuum. Assuming
that $\phi$ has only a mass term in the bulk, one  easily finds that
Fourier modes satisfy
 \be
 \la\phi_n\ra =
 {M_*R\,\la\chi\ra\over 2\pi\left[(\mu R)^2 + \vec{n}^2\right]}~.
 \ee
Therefore, for a large enough $\mu$ we see that many KK modes i.e.
all those  below the threshold for which
 $\vec{n}^2 \ll (\mu R)^2$, get approximately the same vev:
 \be
 \la\phi_n\ra \simeq {M_*R\,\la\chi\ra\over 2\pi(\mu R)^2}~.
 \ee
Notice, however, that each individual vev in this case is quite
suppressed since $\la\chi\ra<M_*$. Nevertheless their accumulated
effect, due to its large number, can easily overcome such a
suppression when contributing to KK graviton decay since there
are about $(\mu_{KK} R)^2$ of such modes.

In the presence of the vevs for many KK modes of the scalar
field, there would a be residual vacuum energy in the five
dimensional theory. If it is too large, it would lead to a
breakdown of the flatness assumption of the fifth direction. It
is therefore important to have an estimate of  the contribution
of $\la\phi_{\vec{n}}\ra$ to the potential $V(\phi(x,y))$. This
can be done by noting that $\la\phi_{\vec{n}}\ra\simeq $ TeV
implies that $\la V(\phi(x,y))\ra\simeq (10^{-7}~GeV)^6$, which
is much less than $M^6_*$ at which value the six dimensional
curvature effect should be important. We therefore believe that
the assumption of flatness of the two extra dimensions is not
unreasonable.

Another radical and interesting alternative is to have a lattice of
defects (e.g. vortices or domain walls) of lattice size $\approx l$
filling up the extra two dimensions and have the KK modes of gravitons
live on lattice sites. These defects can cause a breakdown of the extra-D
momentum and allow the higher KK modes of gravitons to decay, leading to
phenomena similar to what is discussed in the paper. This idea is
presently under study by one of the authors (S.N.).

\section{Conclusions}

In conclusion, we have raised the possibility of a new class of extra
dimensional models
where the translation invariance along the fifth and sixth dimensions is
broken by the vevs of the KK modes of a bulk scalar field. This allows the
higher KK modes of
the gravitons to undergo fast decay to all the lower lying KK modes of the
graviton. The decay branching ratio of the higher KK modes of the
graviton to standard model particles such as photons is then reduced
by a significant amount with very interesting cosmological and
astrophysical
implications. For example, it now allows for the size of the extra two
dimensions to be in millimeter range (with the higher dimensional gravity
scale in TeV range as required to solve the gauge hierarchy
problem) without conflicting with several astrophysical and cosmological
bounds. At present, we do not see how this mechanism will be of help
in relaxing the bounds from the supernova emission rates. Nonetheless,
this is a new and interesting possibility that in our opinion deserves
consideration. 
This possibility should also have implications for collider
signatures of the TeV scale gravity models, that will be the subject of a
future investigation.

\acknowledgements

A.P.L. would like to thank the Particle Theory group of the
University of Maryland for warm hospitality. The works of R. N. M. and
partly of A.P.L. are supported by the National Science Foundation Grant
No. PHY-0099544. We thank T. Jacobson and M. Luty for discussions.


\end{document}